\documentclass[]{pasj01}

\Received{}
\Accepted{}
 
 \usepackage[T1]{fontenc} 

\begin{document} 

\title{ 
\LETTERLABEL 
ALMA Observations of PSR B1259-63/LS 2883 in an Inactive Period: Variable Circumstellar Disk?}

\author{Yutaka \textsc{Fujita}\altaffilmark{1}$^{*}$}%
\altaffiltext{1}{Department of Physics, Graduate School of Science,
Tokyo Metropolitan University,
1-1 Minami-Osawa, Hachioji-shi, Tokyo 192-0397, Japan}
\email{y-fujita@tmu.ac.jp}

\author{Hiroshi \textsc{Nagai}\altaffilmark{2}}
\altaffiltext{2}{National Astronomical Observatory of Japan, 2-21-1 Osawa, Mitaka, Tokyo 181-8588, Japan}

\author{Takuya \textsc{Akahori}\altaffilmark{2}}

\author{Akiko \textsc{Kawachi}\altaffilmark{3}}
\altaffiltext{3}{Department of Physics, School of Science, Tokai
University, Kitakaname, Hiratsuka, Kanagawa 259-1292, Japan}

\author{Atsuo T. \textsc{Okazaki}\altaffilmark{4}}
\altaffiltext{4}{Faculty of Engineering, Hokkai-Gakuen University, 1-40, 4-chome, Asahi-machi, Toyohira-ku, Sapporo 062-8605,
Japan}



\maketitle

\begin{abstract}
We report Atacama Large Millimeter/submillimeter Array (ALMA)
observations of the gamma-ray binary system containing the pulsar
PSR~B1259--63 orbiting around a massive star LS~2883 in an inactive
period between the 2017 and 2021 periastron passages. We detected radio
continuum emission from the binary system at 97~GHz (Band~3) and 343~GHz
(Band~7). Compared with our previous ALMA observations performed soon
after the 2017 periastron passage, the fluxes have decreased by an
factor of six at 97~GHz and two at 343~GHz. The flux at 343~GHz is large
relative to that at 97~GHz and appears to be thermal emission from the
circumstellar disk around LS~2883. The decrease of the 343~GHz
flux may indicate that the disk has expanded and become partially
optically-thin since the disk is no longer affected by pulsar winds.
The flux at 97~GHz is consistent with that expected from the pulsed
emission from the pulsar, which indicates that the unpulsed emission
that had been produced through pulsar-disk or pulsar-stellar wind
interaction has disappeared. The image of the system is consistent with
a point source and shows no sign of ejecta.
\end{abstract}

\section{Introduction}

High-mass gamma-ray binaries consist of a massive B- or O-type star and
a compact object (a neutron star or a black hole).  So far, only seven
gamma-ray binary systems have been confirmed. PSR~B1259-63/LS~2883
(B1259 hereafter) is the first of these binaries whose compact object
has been identified as a radio pulsar or a neutron star. In this system,
the pulsar PSR B1259-63 is orbiting around the rapidly-rotating, late
Oe-type or early Be-type massive ($\gtrsim 10\: M_\odot$) companion star
LS~2883 with a period of 1236.7~d and an orbital eccentricity of 0.87
\citep{2011ApJ...732L..11N,2014MNRAS.437.3255S}. The spin period and the
characteristic age of the pulsar are $P=47.76$~ms and
$P/(2\dot{P})=330$~kyr, respectively
\citep{1992ApJ...387L..37J,1994MNRAS.268..430J}. The distance to the
system is $2.6^{+0.4}_{-0.3}$~kpc \citep{2018MNRAS.479.4849M}.  Around
periastron, the pulsed emission becomes undetectable for $\sim 35$~days,
which indicates that the pulsar is eclipsed by the circumstellar disk of
LS~2883 \citep{1996MNRAS.279.1026J}. On the other hand, B1259 emits
unpulsed radio and high-energy emissions, which are probably produced by
interaction between the pulsar and the circumstellar disk and/or stellar
winds \citep{1996MNRAS.279.1026J,2011ApJ...736L..11A}. In fact, double
peaked light curves have been observed in radio and X-rays around
periastron, each of which reflects disk-crossing of the pulsar
\citep{2002MNRAS.336.1201C,2005MNRAS.358.1069J}. While the radio
emission appears to be synchrotron emission from electrons, the X-ray
emission may be synchrotron emission or inverse Compton (IC) scattering
\citep{2006MNRAS.367.1201C,2009MNRAS.397.2123C,2009ApJ...698..911U,2011MNRAS.417..532P,2015MNRAS.454.1358C}.

Close to periastron passages, TeV gamma-ray emission has been detected
with the High Energy Stereoscopic System (HESS)
\citep{2005A&A...442....1A,2009A&A...507..389A}. In an IC scenario,
high-energy electrons upscatter soft UV photons stemming from the
stellar and/or disk radiation field into the gamma-ray regime
\citep{1996A&AS..120C.221T,1999APh....10...31K,2011MNRAS.416.1067K}.
The generation of TeV gamma-rays within a hadronic scenario is another
explanation. In this model, the dense circumstellar disk is an ideal
source of target material of ultrarelativistic particles that could
produce pions and hence TeV gamma-rays
\citep{2004ApJ...607..949K,2007Ap&SS.309..253N}. GeV gamma-ray flares
have been observed with the Fermi Gamma-ray Space Telescope after the
passage of the pulsar through the circumstellar disk
\citep{2011ApJ...736L..11A,2015ApJ...811...68C,2018ApJ...862..165T,2018ApJ...863...27J}.
The flares seem to be triggered by the disruption of the disk
\citep{2015MNRAS.454.1358C}.

\begin{figure}
 \begin{center}
  \includegraphics[width=8cm]{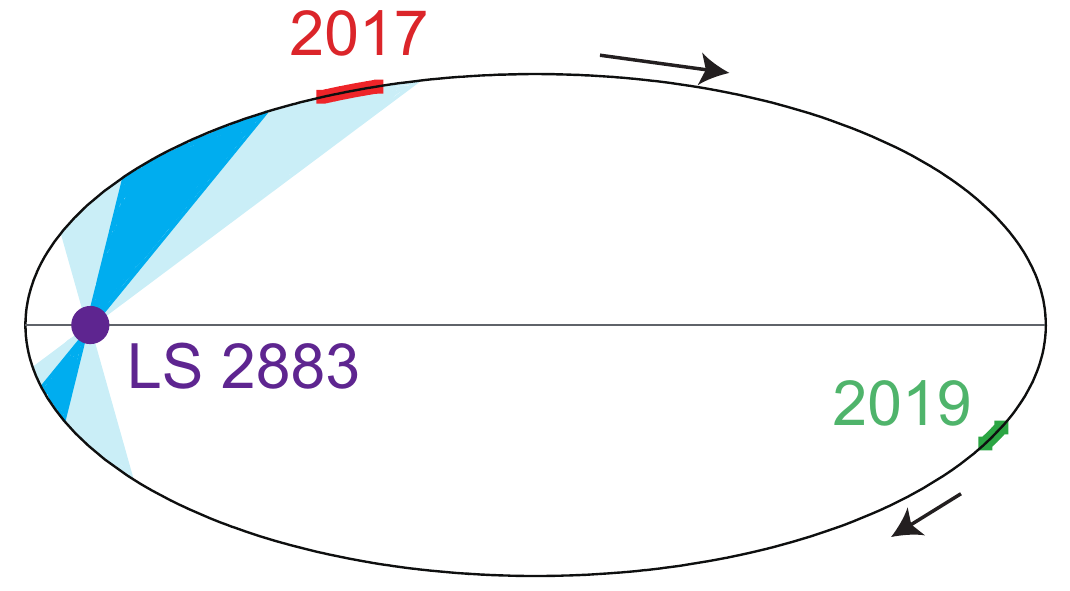} 
 \end{center}
\caption{Schematic representation of B1259 with the locations of the
pulsar during the 2017 (red) and 2019 (green) observations with
ALMA. Shaded area shows the geometry of the disk inferred from the X-ray
data \citep{2006MNRAS.367.1201C}. Darker and lighter shaded regions
correspond to one and two half-opening angles of the disc,
respectively.}\label{fig:orbit}
\end{figure}

Until recently, there had been no observations at $\nu\sim
100$--$300$~GHz. However, we observed B1259 with Atacama Large
Millimeter/submillimeter Array (ALMA) after the 2017 periastron passage
and detected it in the millimeter and submillimeter wavelengths for the
first time (\cite{2019PASJ...71L...3F}; hereafter Paper~I). Our
observations were made soon after the disk-crossing of the pulsar
(figure~\ref{fig:orbit}). At Band 3 (97~GHz), the flux 84 days after the
periastron was almost identical to that 71 days after the periastron.
Compared with a GeV gamma-ray light curve, we concluded that the
emission at 97~GHz is not related to the gamma-ray flares. The 97~GHz
fluxes were consistent with an extrapolation of the radio spectrum at
lower frequencies ($\lesssim 10$~GHz). Thus, we speculated that it is
the unpulsed synchrotron emission from electrons accelerated when the
pulsar passed through the circumstellar disk. The flux at Band 7 (343
GHz) 69 days after the periastron was significantly larger that the
extrapolation of the radio spectrum at lower frequencies. We argued 
that this emission is thermal one coming from the circumstellar disk
around LS~2883.

In this letter, we report the results of our new observations of B1259
with ALMA around apastron (figure~\ref{fig:orbit}). Previous
observations at low-frequencies ($\nu\lesssim 10$~GHz) have indicated
that unpulsed synchrotron emission disappears in this period
\citep{1999MNRAS.302..277J,2005MNRAS.358.1069J}. If the 97~GHz flux is
associated with that emission, it should significantly decrease or
disappear accordingly. On the other hand, if the 343~GHz emission
actually comes from the disk, changes of the flux implies a long-term
variability of the disk.

\section{Observations and Results}
\label{sect:obs} 

The observations were carried out in ALMA Band 3 and Band 7 in 2019
November.  The data were taken in Time Division Mode (TDM) centered at
the frequency of 97~GHz and 343~GHz at Band 3 and Band 7,
respectively. The data were processed using CASA 5.6.1-8 and ALMA
Pipeline-CASA56-P1-B in a standard manner.  Number of antennas used for
the observation, total on-source time, bandpass calibrator, and gain
calibrator are summarized in table \ref{tab:obs}.  Flux scaling was
derived on the bandpass calibrator using the flux information provided
by the Joint ALMA Observatory (JAO). The observation condition in terms
of the weather condition and precipitable water vapor (PWV) was normal
in both Band 3 and 7, but the phase RMS per baseline visibility for the
second observation in Band 7 (Nov. 27) was at most 1.5 times larger than
that for the first observation (Nov. 7) according to the Quality
Assurance (QA) 0 report.  The image deconvolution was done using the
CASA task \texttt{tclean} with the multifrequency synthesis. Since the
second observation of Band 7 was carried out with a longer baseline
configuration, we applied uv-taper option to have a beam size similar to
the first observation. Table \ref{tab:flux} summarizes angular
resolution and image rms.

\begin{figure*}
\begin{center}
\includegraphics[width=18cm]{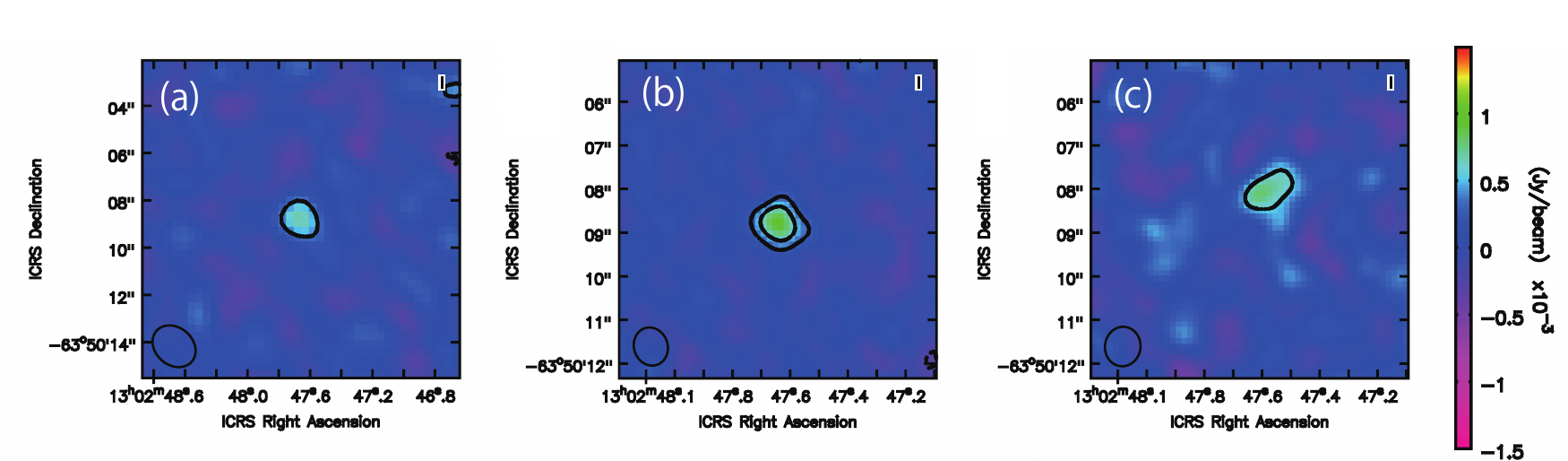}
\end{center}
\caption{(a) Band 3 image of B1259 on 2019 November 2 ($t_p=771$~d). (b) Band 7 image of B1259 on 2019 November 7 ($t_p=776$~d). (c) Band 7 image of B1259 on 2019 November 27 ($t_p=796$~d).  Contours are plotted at the level of $n\times3\sigma$ ($n=1, 2$) where $\sigma$ is image RMS summarized in Table \ref{tab:flux}.}\label{fig:target}
\end{figure*}

Figure~\ref{fig:target} shows the total intensity image of
B1259. Significant emission is detected both in Band 3 and 7.  The
source structure is point-like (figures~\ref{fig:target}a and b) but
somewhat slightly extended in the second observation of Band 7
(figure~\ref{fig:target}c). We consider that the extended structure is
not real but caused by a phase calibration error since the emission is
shifted from the phase center owing to the phase calibration
error and a similar level of shift and image distortion is seen in a
calibrator J1337-6509, which was observed as the check source (section
10 in ALMA Technical Handbook: Remijan et al. 2020\footnote{https://almascience.nrao.edu/documents-and-tools/cycle7/alma-technical-handbook/view}). Because of the poor
phase calibration, the second observation data is likely affected by
coherence loss. In fact, the peak intensity appears to be only about
70\% of that for the first observation.

The obtained fluxes are shown in table~\ref{tab:flux}. We also show
those for our 2017 observations for comparison. The most recent
periastron passage of the pulsar occurred on MJD~58018.143 (UTC 2017
September 22 03:25:55.2). We refer to the time of the passage as $t_p$.
For the flux estimation, we made Gaussian model fitting to the images
using the CASA task \texttt{imfit} and measured the integrated flux
density.  For the second observation of Band 7 ($t_p=796$~d), the
structure is distorted (figure~\ref{fig:target}c) and thus the Gaussian
is not a good representation of the source structure.  Besides, the flux
is likely underestimated by coherence loss due to the poor phase
calibration.  Thus, we show a peak intensity for the second observation
of Band 7 as a lower limit of flux density. In the following sections,
we do not discuss the second observation of Band~7 
($t_p=796$~d).

\begin{table*}
\tbl{Summary of 2019 observations}{
\begin{tabular}{ccccccc}
\hline
Band & Center frq. & Date & $N_{\rm ant}$$^{a}$ & On-source time$^{b}$ & Bandpass/Flux$^{c}$ & Gain$^{d}$ \\ \hline
3 & 97~GHz & November 2, 2019 & 43 & 5.5 min & J1617-5848 & J1322-6532 \\
7 & 343~GHz & November 7, 2019 & 44 & 5 min & J1107-4449 & J1254-6111 \\
7 & 343~GHz & November 27, 2019 & 41 & 5 min &  J1337-1257 & J1254-6111 \\
\hline
\end{tabular}}
\label{tab:obs}
\begin{tabnote}
\footnotemark[$a$] Number of antennas used for observation. \footnotemark[$b$] Total integration time of the target source. \footnotemark[$c$] Bandpass and flux calibrator name. \footnotemark[$d$] Gain calibrator name.
\end{tabnote}
\end{table*}

\begin{table*}
  \tbl{Angular resolution, image rms, and observed fluxes for the ALMA
  observations in 2017 and 2019}{
  \begin{tabular}{cccccc}
 \hline
Band & Date & Day & Beam Shape & Image RMS  & Observed Flux  \\ 
     & & (from $t_{\rm p}$) &  & ($\mu$Jy~beam$^{-1}$) & (mJy)  \\ \hline   
 3 (97~GHz) & Dec. 2, 2017 & +71 & $0.35"\times0.21"$ at $78^\circ$ & 41 & $1.1\pm 0.1$ \\
 3 (97~GHz) & Dec. 15, 2017 & +84 & $0.42"\times0.36"$ at $-52^\circ$ & 36 & $0.97\pm 0.09$\\
 7 (343~GHz) & Nov. 30, 2017 & +69 & $0.056"\times0.043"$ at $-8^\circ$ & 87 & $2.3\pm 0.4$\\ 
 3 (97~GHz) & Nov. 2, 2019 & +771 & $1.98"\times1.58"$ at $49^{\circ}$ & 40 & $0.19\pm 0.04$\\
 7 (343~GHz) & Nov. 7, 2019 & +776 & $0.88"\times0.77"$ at $18^{\circ}$ & 90 & $1.12\pm 0.13$\\
 7 (343~GHz) & Nov. 27, 2019 & +796 & $0.91"\times0.81"$ at $-7^{\circ}$ & 179 & $>0.8$ \\
\hline
  \end{tabular}}\label{tab:flux}
\end{table*}

\section{Discussion}

\subsection{Comparison with 2017 observations}

\begin{figure}
 \begin{center}
  \includegraphics[width=8cm]{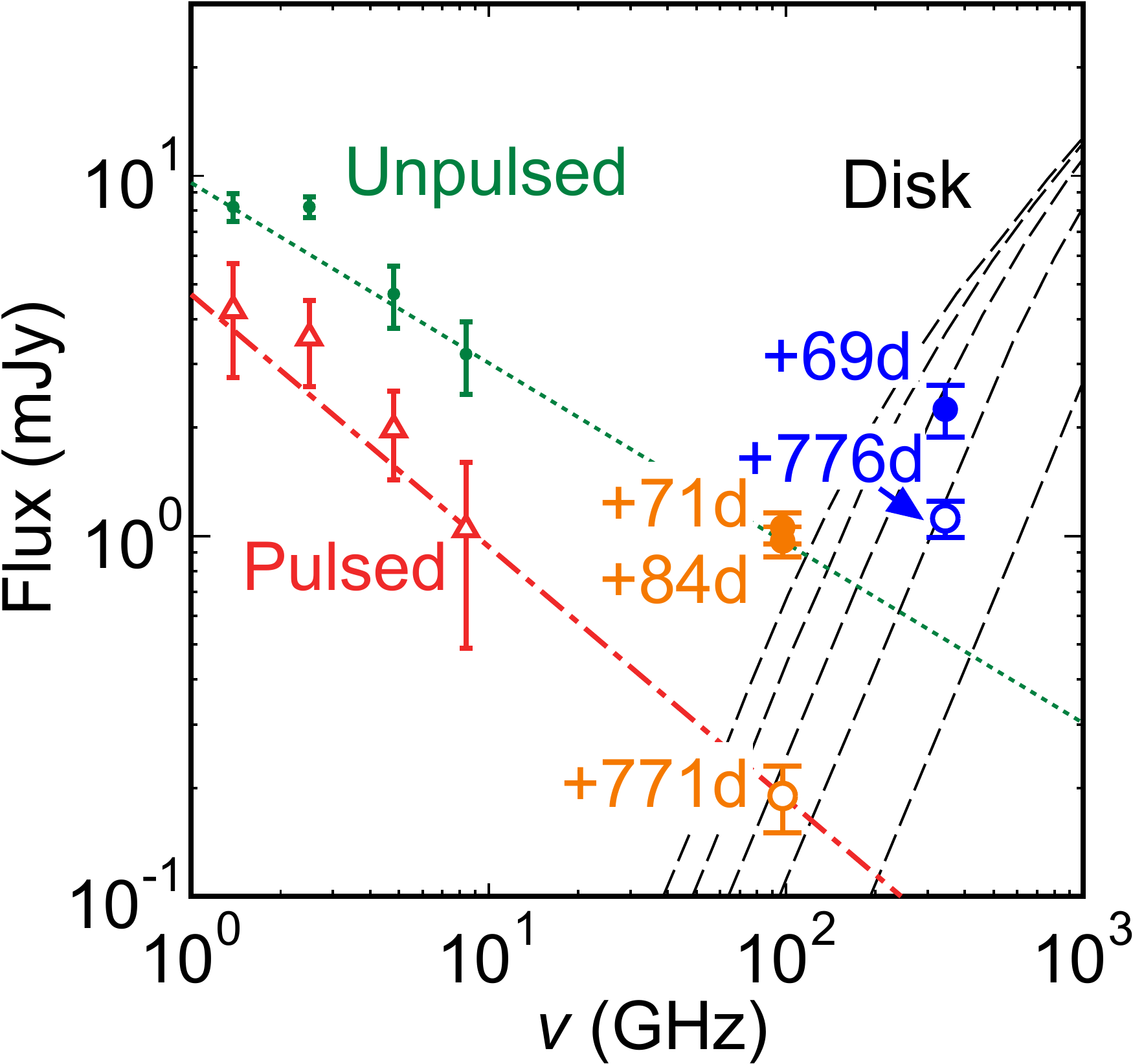} 
 \end{center}
\caption{Radio fluxes of B1259. Our new ALMA observations in 2019 are
shown by open circles at 97~GHz (orange) and 343~GHz (blue). The labels
are the days since $t_p$. Our previous ALMA observations in 2017 are
also shown by filled circles at 97 (orange) and 343~GHz (blue).
Unpulsed fluxes obtained with ATCA at $t_p + 64.2$~d at the 2004 passage
are shown by green dots \citep{2005MNRAS.358.1069J}.  The dotted green
line is a spectrum represented by $S_\nu \propto \nu^{-0.5}$. The
normalization is set so that the line passes 2017 observations at
97~GHz. Averaged pulsed fluxes obtained with ATCA before and after the
2004 passage are shown by red open triangles
\citep{2005MNRAS.358.1069J}. The associated red bars are the standard
deviations of the data at individual frequencies. The dashed-dotted red
line is a spectrum represented by $S_\nu \propto \nu^{-0.7}$. The
normalization is set so that the line passes the new ALMA observation at
97~GHz. Dashed black lines show the infrared emission from the
circumstellar disk of LS~2883 for different disk sizes predicted by
\citet{2011MNRAS.412.1721V}. The disk sizes are (from the bottom to the
top) 10, 20, 30, 40 and 50~$R_*$. }\label{fig:radio}
\end{figure}

In figure~\ref{fig:radio}, the radio fluxes obtained through our new
ALMA observations in 2019 are shown by open circles. For comparison, we
also present the results of our previous ALMA observations in 2017 by
filled circles. The figure indicates that the fluxes in 2019 have
decreased since our previous observations in 2017 especially at
97~GHz. In Paper~I, we compared our 2017 observations with ATCA
observations at low-frequencies at the 2004 periastron
passage\footnote{Unfortunately, the fluxes at the low-frequencies for
the 2017 periastron passage have not been reported.} (green dots in
figure~\ref{fig:radio}). While fluxes at 97~GHz ($t_p+71$~d and
$t_p+84$~d) are in line with an extrapolation of the unpulsed ATCA
fluxes at a similar orbital phase (green dotted line), that at 343~GHz
($t_p+69$~d) is clearly above the extrapolation. We interpreted that
synchrotron emission generated through the interaction between the
pulsar wind and the circumstellar disk is responsible for the 97~GHz
flux. On the other hand, we speculated that the flux at 343~GHz is
associated with the radiation from the circumstellar disk. By comparing
the results with a theoretical model by \citet{2011MNRAS.412.1721V}, we
estimated that the disk size is $\sim 30\: R_*$ in 2017 ($t_p+69$~d in
figure~\ref{fig:radio}), where $R_*$ is the radius of LS~2883. Following
\citet{2011MNRAS.412.1721V}, we here assume that $R_*\sim 6\: R_\odot$
\citep{1994MNRAS.268..430J}, although some recent observations suggest
that $R_*\sim 9\: R_\odot$ \citep{2011ApJ...732L..11N}.

\begin{figure}
 \begin{center}
  \includegraphics[width=8cm]{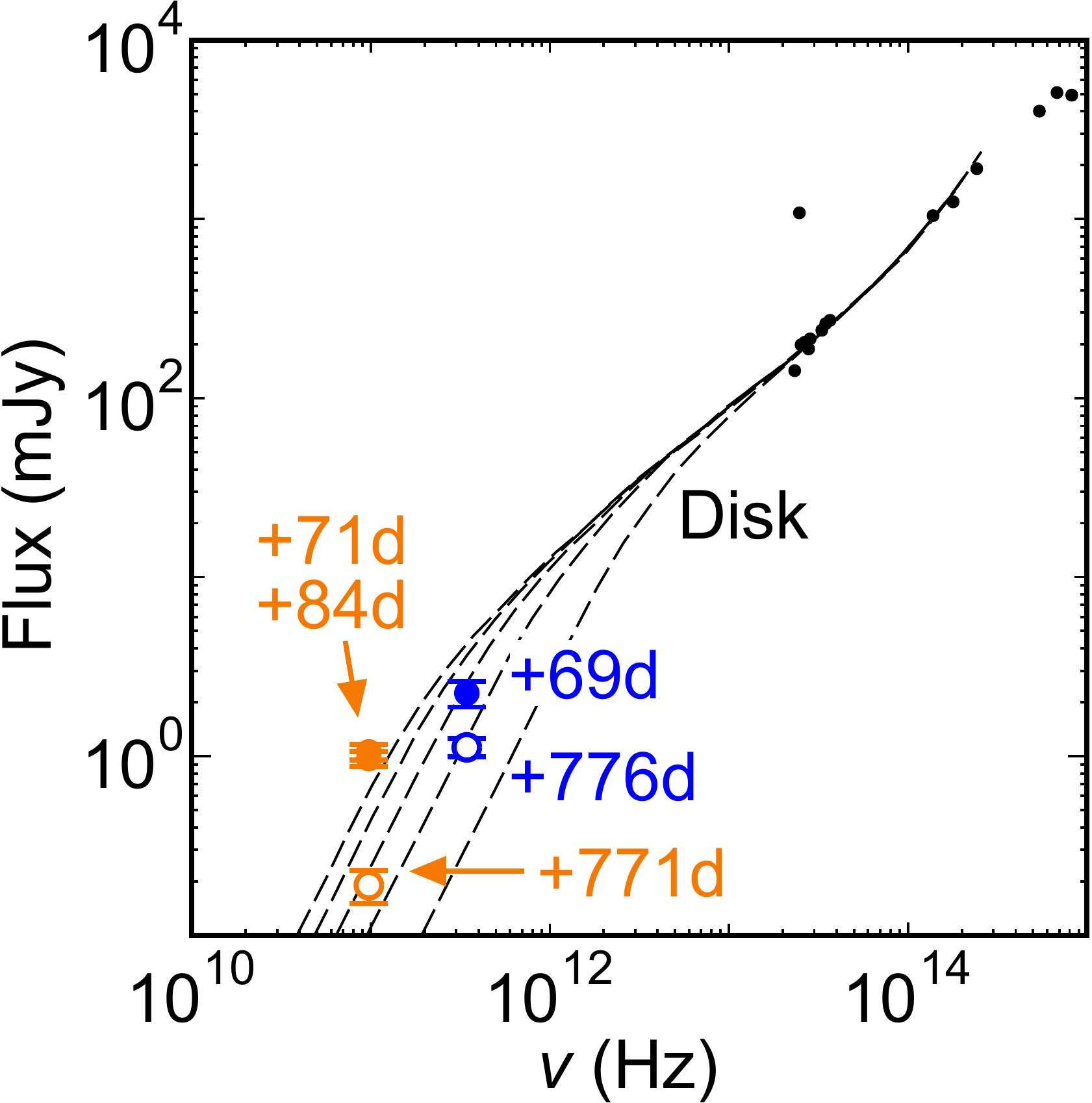} 
 \end{center}
\caption{Radio fluxes of B1259. Our new ALMA observations in 2019
are shown by open circles (Band~3 or 97~GHz (orange) and Band~7 or
343~GHz (blue)). The labels are the days since $t_p$. Our previous ALMA
observations in 2017 are also shown by filled circles at 97 and 343~GHz.
Dashed black lines show the infrared emission from the Be star's
circumstellar disk for different disk radii predicted by
\citet{2011MNRAS.412.1721V}. The disk radii are (from the bottom to the
top) 10, 20, 30, 40 and 50~$R_*$. The observational data (black dots) are those compiled by \authorcite{2012MNRAS.426.3135V} (\yearcite{2012MNRAS.426.3135V}; see also \cite{1989A&AS...78..203W,2006AJ....131.1163S,2001AJ....121.2819P,2011ApJ...732L..11N}).}\label{fig:ir}
\end{figure}

In figure~\ref{fig:ir}, we show the disk spectra predicted by
\citet{2011MNRAS.412.1721V} up to the optical band.  They are
insensitive to the disk radius at $\nu\gtrsim 10^{13}$~Hz, because the
disk is only partially optically-thick and most emissions come from the
inner disk \citep{1986A&A...162..121W}. On the other hand, the spectra
are sensitive to the disk size in the ALMA band ($\nu\sim
100$--300~GHz), because the whole disk is optically-thick
\citep{1986A&A...162..121W}. In fact, at $\nu\lesssim 300$~GHz, the
dashed lines follow $\propto \nu^2$, or black-body radiation.  We plot
the results of our ALMA observations in figure~\ref{fig:ir}. The
spectral index of the observed fluxes between 97~GHz and 343~GHz is
$\alpha =0.7^{+0.2}_{-0.3}$ for the 2017 observations and $\alpha =
1.4^{+0.2}_{-0.1}$ for the 2019 observations, where the index is defined
as $S_\nu\propto \nu^\alpha$. Since the indices are smaller than
$\alpha=2$, the whole emission in the ALMA bands cannot be explained by
pure black-body disk radiation.

\subsection{Origin of the radio emissions}

According to our interpretation in Paper~I, the decrease of the flux at
343~GHz in 2019 could be interpreted that the disk radius has decreased
down to $\sim 20\: R_*$ in 2019 ($t_p+776$~d in
figure~\ref{fig:radio}). If this is the case, the expected disk flux at
97~GHz is $\sim 0.1$~mJy, which is smaller than the flux of 0.19~mJy
measured in 2019 ($t_p+771$~d in figure~\ref{fig:radio}). One possible
explanation is that it is the pulsed emission from the pulsar. In
figure~\ref{fig:radio}, red open triangles are the averaged pulsed
fluxes obtained with ATCA during the period between $t_p - 21.5$~d and
$t_p + 186.1$~d at the 2004 periastron passage
\citep{2005MNRAS.358.1069J}. The flux at 97~GHz in 2019 is on the
extrapolation of the pulsed fluxes (dashed-dotted red line in
figure~\ref{fig:radio}). This suggests that the unpulsed component
(green dotted line) had fallen below the pulsed
component. Unfortunately, we cannot uncover the pulse (47.76~ms) from
the ALMA data, because they are integrated at every two seconds.

If the decrease of the 343~GHz flux actually reflects that of the disk
size, it may mean that the disk evolves on a time-scale of a
year. Numerical simulations have shown that the disk is disrupted when
the pulsar approaches the disk around periastron
\citep{2011PASJ...63..893O,2012ApJ...750...70T}. Non-thermal X-ray
emissions are produced through the interaction between the pulsar wind
and the disk around this period \citep{2012ApJ...750...70T}. However,
these simulations suggest that the disk should recover after the pulsar
moves away from the disk. Thus, if the disk destruction by the
pulsar is only the cause of the disk variation, the flux from the disk
in 2019 is expected to increase from that in 2017, which is opposite to
our observational results.

Assuming that the the 343~GHz flux is the black-body radiation
from the disk, another possibility is that the disk has {\it expanded}
but the optically-thick region has shrunk.  This is because when the
pulsar moves away from the Be star towards apastron, the outer part of
the disk, which was strongly compressed and radio optically thick near
periastron, is likely to expand and become optically thin due to the
decrease of the density. If this is the case, the luminosity of the
disk should decrease, because it reflects the area of the optically-thick
region. Since the geometry of this binary system is complicated,
numerical simulations would be required to confirm this speculation.

\subsection{X-ray ejecta}

{\it Chandra} X-ray observations have revealed the existence of
high-speed ejecta that appeared as an extended X-ray structure (clump)
moving away from B1259
\citep{2011ApJ...730....2P,2014ApJ...784..124K,2015ApJ...806..192P}. The
clump is ejected at every binary cycle near periastron passage
\citep{2019ApJ...882...74H}. Although we have checked the ALMA images
for both 2017 and 2019 observations, they are consistent with a point
source (see section~\ref{sect:obs} and figure~\ref{fig:target} for 2019
observations). The upper-limit of the image size is given by the beam
shapes (table~\ref{tab:flux}).

\section{Conclusions}

We have reported the results of ALMA observations of the pulsar-massive
star binary PSR B1259-63/LS~2883 at 97 and 343~GHz. The observations
were made in 2019 around the apastron after the 2017 periastron
passage. We detected emissions from the system both at 97~GHz and
343~GHz. However, the fluxes have decreased by a factor of six at 97~GHz
and two at 343~GHz since our observations that were made soon after the
2017 periastron passage. We argued that while the emission at 343~GHz is
a thermal emission from the circumstellar disk around LS~2883, that at
97~GHz is non-thermal pulsed emission from the pulsar. In this
scenario, the decrease of the 343~GHz flux may indicate that the radio
optical depths of the Be disk has significantly decreased since the 2017
periastron.  This could happen if the disk density, particularly in the
outer region, has decreased due to the expansion of the disk, because it
is no longer affected by the pulsar wind ram-pressure near apastron.
This speculation could be studied by numerical simulations in the
future.  The fall of flux at 97~GHz indicates that the synchrotron
radiation observed soon after the 2017 periastron had disappeared.  We
also checked the image of the system and found that it is consistent
with a point source. This means that in the radio band there is no sign
of ejecta that had been observed in X-rays.


\begin{ack}
This work was supported by MEXT KAKENHI No. 18K03647 (Y.F.).  This
paper makes use of the following ALMA data:
ADS/JAO.ALMA\#2019.1.00320.S. and 2017.1.01188.S. ALMA is a partnership
of ESO (representing its member states), NSF (USA) and NINS (Japan),
together with NRC (Canada), MOST and ASIAA (Taiwan), and KASI (Republic
of Korea), in cooperation with the Republic of Chile. The Joint ALMA
Observatory is operated by ESO, AUI/NRAO and NAOJ.
\end{ack}


\end{document}